\documentclass[conference]{IEEEtran}
\ifCLASSINFOpdf
\else
\fi
\usepackage{amsmath,dsfont,bbm,epsfig,amssymb,amsfonts,amstext,verbatim,amsopn,cite,subfigure,multirow,multicol,lipsum,xfrac}
\usepackage{amsthm}
\usepackage{mathtools,amsthm}
\usepackage{perpage}
\usepackage{balance}
\usepackage{url}
\usepackage{amsfonts}
\usepackage{epsfig}
\usepackage[font={small}]{caption}
\usepackage{psfrag}	
\usepackage{etoolbox}
\usepackage{algorithmicx}
\usepackage[Algorithm,ruled]{algorithm}
\usepackage{algpseudocode}
\usepackage{pifont}
\usepackage[utf8]{inputenc}
\usepackage[T1]{fontenc}  
\usepackage[nolist]{acronym}
\MakePerPage{footnote}
\usepackage{paralist}
\usepackage{enumitem}
\usepackage{bbm}
\usepackage[process=auto]{pstool}
\usepackage{tikz}
\usetikzlibrary{shapes,arrows}
\newcommand{\setX}{\mathbbmss{X}}

\newcommand{\setR}{\mathbbmss{R}}

\newcommand{\setW}{\mathbbmss{W}}
\newcommand{\setJ}{\mathbbmss{J}}
\newcommand{\setZ}{\mathbbmss{Z}}

\newcommand{\setC}{\mathbbmss{C}}

\newcommand{\setT}{\mathbbmss{T}}

\newcommand{\rmP}{\mathrm{P}}

\newcommand{\rmp}{\mathrm{p}}

\newcommand{\rmR}{\mathrm{R}}

\newcommand{\rmG}{\mathrm{G}}
\newcommand{\rmQ}{\mathrm{Q}}

\newcommand{\her}{\mathsf{H}}

\newcommand{\sfM}{\mathsf{M}}

\newcommand{\sfD}{\mathsf{D}}

\newcommand{\sfK}{\mathsf{K}}
\newcommand{\sfp}{\mathsf{p}}
\newcommand{\sfa}{\eta}

\newcommand{\maf}{\mathcal{F}}

\newcommand{\mae}{\mathcal{E}}
\newcommand{\maz}{\mathcal{Z}}
\newcommand{\mas}{\mathcal{S}}

\newcommand{\mg}{\mathcal{G}}
\newcommand{\mai}{\mathcal{I}}

\newcommand{\bx}{{\boldsymbol{x}}}

\newcommand{\xx}{\mathrm{x}}

\newcommand{\set}[1]{\left\lbrace#1\right\rbrace}

\newcommand{\bs}{{\boldsymbol{s}}}

\newcommand{\bv}{{\boldsymbol{v}}}

\newcommand{\dif}{\mathrm{d}}

\newcommand{\mI}{\mathbf{I}}
\newcommand{\mone}{\mathbf{1}}
\newcommand{\mJ}{\mathbf{J}}

\newcommand{\mQ}{\mathbf{Q}}

\newcommand{\mU}{\mathbf{U}}
\newcommand{\mD}{\mathbf{D}}
\newcommand{\mM}{\mathbf{M}}

\newcommand{\mT}{\mathbf{T}}
\newcommand{\mH}{\mathbf{H}}
\newcommand{\mP}{\mathbf{P}}

\newcommand{\md}{\mathrm{D}}
\newcommand{\E}{\mathsf{E}\hspace{.5mm}}

\newcommand{\rs}{{\mathsf{rs}}}

\newcommand{\papr}{{\mathsf{PAPR}}}

\newcommand{\norm}[1]{\lVert #1 \rVert}
\newcommand{\re}[1]{\mathsf{Re}\left\lbrace #1 \right\rbrace}
\newcommand{\img}[1]{\mathsf{Img}\left\lbrace #1 \right\rbrace}

\newcommand{\abs}[1]{\lvert #1 \rvert}

\newtheoremstyle{mystyle}
  {}
  {}
  {}
  {}
  {\bfseries}
  {:}
  { }
  {}

\theoremstyle{mystyle}

\newtheorem{definition}{Definition}
\newtheorem{proposition}{Proposition}
\newtheorem{remark}{Remark}
\newtheorem*{prf}{Proof}
\newcounter{bar}

\begin{document}
%
\title{Nonlinear Precoders for Massive MIMO Systems with General Constraints}

\author{
\IEEEauthorblockN{
Ali Bereyhi, 
Mohammad Ali Sedaghat, 
Saba Asaad, 
Ralf R. M\"uller
}
\IEEEauthorblockA{
Institute for Digital Communications (IDC), Friedrich-Alexander Universit\"at Erlangen-N\"urnberg (FAU)\\
ali.bereyhi@fau.de, mohammad.sedaghat@fau.de, saba.asaad@fau.de, ralf.r.mueller@fau.de 
\thanks{This work was supported by the German Research Foundation, Deutsche Forschungsgemeinschaft (DFG), under Grant No. MU 3735/2-1.}
}
}


%

\IEEEspecialpapernotice{(Invited Paper)}

\IEEEoverridecommandlockouts

\maketitle

\begin{abstract}
We introduce a class of nonlinear least square error precoders with a general penalty function for multiuser massive MIMO systems. The generality of the penalty function allows us to consider several hardware limitations including transmitters with a predefined constellation and restricted number of active antennas. The large-system performance is then investigated via the replica method under the assumption of replica symmetry. It is shown that the least square precoders exhibit the ``marginal decoupling property'' meaning that the marginal distributions of all precoded symbols converge to a deterministic distribution. As a result, the asymptotic performance of the precoders is described by an equivalent single-user system. To address some applications of the results, we further study the asymptotic performance of the precoders when both the peak-to-average power~ratio~and number of active transmit antennas are constrained. Our~numerical~investigations show that for a desired distortion at the receiver side, proposed forms of the least square precoders need to employ around $35\%$ fewer number of active antennas compared to cases with random transmit antenna selection.\\
\end{abstract}

\IEEEpeerreviewmaketitle

\section{Introduction}
The concept of massive \ac{mimo} systems has been considered to be a key solution to vast demands for higher performance gains such as spectral efficiency, rate reliability, and energy efficiency \cite{marzetta2010noncooperative}. These benefits are obtained along with hardware challenges arising due to the tremendous number of antennas. Some of these issues, such as having a large antenna array within a relatively small physical platform, are effectively addressed by utilizing the millimeter wave spectrum \cite{rappaport2013millimeter}; however, some other issues such as overall \ac{rf} cost and energy efficiency are still remained~as~bottlenecks~and need to be overcome by effective design of~system~modules.

In this paper, our concentration is on the downlink scenario. The module which plays a key role in this case is the precoder which maps the data signal to a precoded signal in order to compensate the distortion caused by the channel. This leads to a receive signal with low distortion, and therefore, the processing load at the user side reduces significantly. Primary approaches for precoding are the linear schemes which have lower computational complexity, e.g., \ac{mf} and \ac{rzf}. More advanced approaches such as Tomlinson-Harashima \cite{fischer2002space} and vector precoding \cite{peel2005vector}, however, achieve a better performance at the expense of more complexity. Although these schemes take into account the computational complexity of the precoding algorithms, they do not consider limitations imposed to the system by hardware restrictions. As an example, the aforementioned precoders consider the whole complex plane as the set of possible transmit constellation points; the assumption which in practice does not hold, due to the limited \ac{papr} of power amplifiers. The authors in \cite{mohammed2013per} tried to address this drawback partially by proposing the nonlinear ``per-antenna constant envelope precoding'' in which the precoder maps the data signal such that the precoded signal has a constant amplitude. In \cite{sedaghat2017lse}, a class of nonlinear \ac{lse} precoders was introduced where the precoder minimizes the distortion at the receiver side over a general transmission support considering a constraint on the transmit power. The generality of the transmission support enabled the authors in \cite{sedaghat2017lse} to investigate more general scenarios, such as transmitters with restricted \ac{papr}, and discrete transmit constellations. In this paper, we study the nonlinear \ac{lse} precoders considering a general penalty function which enables us to investigate various limitations on the transmitter including constraints on the limited number of \ac{rf}-chains.
\subsection*{Transmit Antenna Selection}
One of the crucial bottlenecks in massive \ac{mimo} systems is the large overall \ac{rf} cost. In the downlink scenario, this is mainly caused by the vast number of \ac{rf} chains connected to transmit antennas. In this case, \ac{tas} can significantly reduce hardware costs without significant performance loss \cite{molisch2005capacity}. Regarding massive \ac{mimo} systems, the optimal \ac{tas} becomes computationally unfeasible, due to exponentially increasing number of searches. Alternatively, suboptimal greedy algorithms with polynomial order of complexity can be employed \cite{gharavi2004fast,gorokhov2003receive}. From the analytical point of view, the study of \ac{tas} algorithms in large-system limits faces more difficulties, and has been addressed in the literature only for some special selection algorithms invoking tools from order statistics; see \cite{li2014energy,asaad2017tas} and references therein.

The limited \ac{rf} cost of the system can be addressed directly at the precoder instead of employing \ac{tas} algorithms. In fact, by constraining the precoded signal to have~a~certain~number of zero entries, the precoder selects a subset of~transmit~antennas while it maps the data signal to the constellation points.
\subsection*{Contributions}
In this paper, a large class of nonlinear \ac{lse} precoders is introduced with a general penalty function which is able to address several transmit limitations. In the large-system limit, we evaluate the performance of the precoder in terms of an equivalent single-user system. Using this asymptotic result, we show that the precoder exhibits an ``asymptotic marginal decoupling property''. This means that the output symbols of the precoder have identical marginal distributions which are equal to the marginal output distribution of~the~equivalent~single-user system. The asymptotic results of this paper are derived via the replica method from statistical mechanics considering \ac{rs}. An introduction to the replica method is given through the asymptotic analyses, in the appendix.
\subsection*{Notation}
We represent scalars, vectors and matrices with non-bold, bold lower case and bold upper case letters, respectively. A $k \times k$ identity matrix is shown by $\mI_k$, and $\mH^{\her}$ indicates the Hermitian of the matrix $\mH$. The set of real and integer numbers are denoted by $\setR$ and $\setZ$, and their corresponding non-negative subsets are shown by superscript $+$; moreover, $\setC$ represents the complex plane. $\norm{\bx}$ denotes the Euclidean norm of the vector $\bx$, and $\norm{\bx}_0$ represents the zero-norm defined as the number of nonzero entries. For a random variable $x$, $\mathrm{p}_x$ represents either the probability mass or probability density function. Moreover, the expectation operator is denoted by $\E$. For sake of compactness, the set of integers $\set{1, \ldots , n}$ is abbreviated as $[1:n]$. Whenever needed, we consider the entries of $\bx$ to be discrete random variables, namely the support $\setX$ to be discrete. Our results, however, are in full generality and hold also for continuous distributions as well.

\section{Problem Formulation}
\label{sec:sys}
The nonlinear \ac{lse} precoding scheme with general penalty, employed in a multiuser \ac{mimo} system, is defined as
\begin{align}
\bx(\bs,\mH)=\arg \min_{\bv \in {\setX^n}} \norm{\mH \bv-\bs}^2 +u(\bv) \label{eq:1}
\end{align}
where $\mH\in \setC^{k\times n}$ and $\bs\in \setC^{k \times 1}$ denote the channel matrix and data vector, and $u(\cdot)$ is a function which decouples, i.e.,
\begin{align}
u(\bv)=\sum_{j=1}^n u(v_j).
\end{align}
We assume $\mH$ to be a random matrix with decomposition 
\begin{align}
\mH^{\her} \mH = \mU \mD \mU^{\her}
\end{align}
where $\mU_{n \times n}$ is a Haar distributed unitary matrix, and $\mD_{n \times n}$ is a diagonal matrix with asymptotic eigenvalue distribution $\rmp_\mD$. The ensemble of $\mH$ encloses a large class of \ac{mimo} channel models including the well-known \ac{iid} Rayleigh fading model. $\bs_{k\times 1}$ is independent of $\mH$ and considered to have \ac{iid} zero-mean complex Gaussian entries with variance $\lambda_s$, i.e., $\bs\sim\mathcal{CN}(\boldsymbol{0}, \lambda_s \mI_{k})$. We let the dimensions of $\mH$ grow large assuming the load factor, defined as $\alpha\coloneqq k/n$, is asymptotically constant in both $k$~and~$n$. For sake of brevity, we drop the dependence of $\bx$ on $\bs$ and $\mH$.

By setting the penalty function $u(\cdot)$ and the support $\setX$ to be of some given forms, the precoder in \eqref{eq:1} reduces to several specific precoders. As examples, let $u(\bv)=\lambda \norm{\bv}^2$; then, 
\begin{enumerate}[label=(\alph*)]
\item by setting $\setX=\setC$, the precoder reduces to the~\ac{rzf}~precoder \cite{peel2005vector} which reads
\begin{align}
\bx= \mH^\her (\mH \mH^\her + \lambda \mI_k)^{-1} \bs.
\end{align}
\item for $\setX$ being a circle in the complex plane, the precoder reduces to the constant envelope precoder considered~in~\cite{mohammed2013per}.
\item when $\setX$ is set to be a general subset of $\setC$, the precoder reduces to the nonlinear \ac{lse} precoder introduced in \cite{sedaghat2017lse}.
\end{enumerate}
The precoder considered in \eqref{eq:1} addresses the above precoding schemes as well as some other techniques which consider different constraints. To study the large-system properties of the precoder, we define the following asymptotic parameters.
\begin{definition}[\bfseries Asymptotic Marginal]
\label{def:margin}
Consider the precoded vector $\bx_{n\times 1}$, and function $f(\cdot)$ defined as
\begin{align}
f(\cdot): \setX \mapsto \setR. \label{eq:4}
\end{align}
Define the marginal of $f(\bx)$ over $\setW(n) \subseteq [1:n]$ as
\begin{align}
\sfM_{f}^{\setW}(\bx;n)\coloneqq \frac{1}{\abs{\setW(n)}} \sum_{w\in\setW(n)} \E f(x_w)
\end{align}
The asymptotic marginal of $f(\bx)$ is then defined to be the large limit of $\sfM^{\setW}_f(\bv;n)$, i.e., $\sfM^{\setW}_f(\bx) \coloneqq \lim\limits_{n\uparrow\infty}\sfM^{\setW}_f(\bx;n)$.
\end{definition}

\begin{definition}[\bfseries Asymptotic Distortion]
\label{def:asy_dist}
For the precoder defined in \eqref{eq:1}, the asymptotic input-output distortion is defined~as
\begin{align}
\sfD\coloneqq\lim_{k\uparrow\infty} \frac{1}{k} \E \norm{\mH \bx-\bs}^2.
\end{align}
where $\bx$ is the precoded vector given in \eqref{eq:1}.
\end{definition}
The asymptotic marginal describes the statistical properties of $\bx$ in the large limit. Moreover, Definition~\ref{def:asy_dist} determines the distortion between the noise-free version of the channel output, i.e., $\mH\bx$, and the data vector. Our goal is to determine the asymptotic distortion as well as the asymptotic marginal of $f(\bx)$ for a given function $f(\cdot)$. To overcome this task, we invoke the replica method. The results then let us investigate several configurations of the precoder which address different criteria such as power constraint, \ac{papr} control and \ac{tas}.

\section{Main Results}
\label{sec:result}
Our main result gives closed-form expression for the asymptotic distortion and marginal of $f(\bx)$ considering a large class of random channel matrices. Before stating the results, let us define the $\rmR$-transform of a given probability distribution.
\begin{definition}[\bfseries $\rmR$-transform]
\label{def:r-trans}
For the random variable $t$ with distribution $\rmp_t$, the Stieltjes transform over the upper complex half plane is given by $\rmG_t(s)= \E (t-s)^{-1}$. Let $\rmG_t^{-1} (\cdot)$ denote the inverse \ac{wrt} composition. Then, the {$\rmR$-transform} of the distribution $\rmp_t$ is defined as $\rmR_t (\omega) = \rmG_t^{-1} (-\omega) - \omega^{-1}$ such that $\lim\limits_{\omega\downarrow 0} \rmR_t (\omega) = \E t$.
\end{definition}
For the random matrix $\mH$ specified in Section~\ref{sec:sys}, the $\rmR$-transform of the Gramian $\mH^\her \mH$ is defined \ac{wrt} the asymptotic eigenvalue distribution $\rmp_\mD$,~and~denoted~by $\rmR_\mD(\cdot)$. Proposition~\ref{thm:1} gives the asymptotics of the precoder in~terms~of~$\rmR_\mD(\cdot)$.
\begin{proposition}[\bfseries \ac{rs} Ansatz]
\label{thm:1}
Consider the nonlinear \ac{lse} precoder defined in Section \ref{sec:sys}, and assume a set of assumptions, including replica continuity and \ac{rs}, holds. Let $s^\rs$ be a zero-mean complex Gaussian random variable with variance $\lambda^\rs$ which for some $\chi$ and $\sfp$ reads as follows
\begin{align}
\lambda^\rs=\left[\rmR_\mD(-\chi)\right]^{-2}\frac{\partial}{\partial \chi} \left[ ( \lambda_s \chi- \sfp ) \rmR_\mD(-\chi) \right].
\end{align}
Moreover, define the random variable $\xx$ to be
\begin{align}
\xx=\arg \min_v \abs{v- s^\rs}^2+ \left[\rmR_\mD(-\chi)\right]^{-1} u(v). \label{eq:snigle}
\end{align}
with $v\in\setX$. Then, the asymptotic marginal of $f(\bx)$ is~given~by
\begin{align}
\sfM^{\setW}_f(\bx)= \E f(\xx),
\end{align}
and the asymptotic distortion is determined as
\begin{align}
\sfD=\lambda_s+\alpha^{-1} \frac{\partial}{\partial \chi} \left[ ( \sfp -\lambda_s \chi ) \chi \rmR_\mD(-\chi) \right]
\end{align}
when $\sfp$ is set to be the average transmit power of the precoder determined as $\sfp\hspace*{-.5mm}=\hspace*{-.5mm} \E \abs{\xx}^2$ and $\chi$ satisfies the fixed~point~equation
\begin{align}
\chi \rmR_\mD (-\chi) &= \frac{1}{\lambda^\rs} \E \re{\xx^* s^\rs} . \label{eq:fix}
\end{align}
\end{proposition}
\begin{prf}
The proof is briefly sketched in the appendix, and the detailed derivations are left for the extended version of the manuscript.
\end{prf}
Proposition \ref{thm:1} determines the asymptotic distortion by solving a set of fixed point equations. It implies moreover that under the set of supposed assumptions, i.e., replica continuity and \ac{rs}
\begin{enumerate}[label=(\alph*)]
\item the asymptotic marginal of $f(\bx)$ does not depend on the index set $\setW(n)$.
\item the asymptotic marginal of $f(\bx)$ is equal to the expected marginal of an equivalent single-user nonlinear \ac{lse} precoder which maps $s^\rs$ to $\xx$ via \eqref{eq:snigle}.
\end{enumerate}
These findings lead us to conclude this property of the~precoder that the output symbols marginally decouple into the outputs of similar single-user nonlinear \ac{lse} precoders. This property is referred to as the ``\ac{rs} marginal decoupling property'' and is deduced directly from Proposition \ref{thm:1}.
\subsection{\ac{rs} Marginal Decoupling Property}
For different classes of estimators, the decoupling property has been investigated in the literature, e.g. \cite{guo2005randomly, rangan2012asymptotic, bereyhi2016rsb}. The main idea of this property is that for a given estimator, the joint distribution of almost any pair of input-output symbols converges to a fixed distribution which is constant in terms of the symbols' index. We show that a marginal version of the decoupling property holds for nonlinear \ac{lse} precoders~as~well.

To illustrate the property, let us denote the marginal distribution of the $j$th symbol of $\bx_{n\times1}$, i.e., $x_j$ for $j\in [1:n]$, by $\rmp^{j(n)}_x$ where the superscript $n$ indicates the dependency on the length of $\bx$. Moreover, we define $\rmp^{j}_x$ to be the asymptotic limit of $\rmp^{j(n)}_x$ meaning that for $t\in\setX$
\begin{align}
\rmp^{j}_x(t)\coloneqq\lim_{n\uparrow\infty} \rmp^{j(n)}_x(t).
\end{align}
The \ac{rs} marginal decoupling property states that, under \ac{rs}, $\rmp^{j}_x$ is constant in $j$ for any $j\in [1:n]$. Therefore, one can consider the precoded symbols to be outputs of copies of a single-user nonlinear \ac{lse} precoder consistent with \eqref{eq:snigle}. More precisely, considering the marginal distributions, the entries of the precoded vector $\bx$ are identically distributed with the distribution of the random variable $\xx$ defined in \eqref{eq:snigle}. We call this random variable ``decoupled precoded symbol'' and denote its marginal distribution~with~$\rmp_\xx$.

\begin{proposition}[\bfseries \ac{rs} Marginal Decoupling Property]
\label{thm:2}
Let the~nonlinear \ac{lse} precoder satisfy the constraints given in Section~\ref{sec:sys}. Then, under some assumptions including the replica continuity and \ac{rs}, the symbol $x_j$, for any $j\in[1:n]$,~converges~in~distribution to the random~variable $\xx$ given in \eqref{eq:snigle}.
\end{proposition}
\begin{prf}
The proof directly follows from Proposition \ref{thm:1}. Let the function $f(\cdot)$ be
\begin{align}
f(x)=\delta(x-t) \label{eq:delta}
\end{align}
and consider the index set $\setJ(n)=[j:j+\zeta n]$. Substituting in Definition \ref{def:margin}, the marginal of $f(\bx)$ reads
\begin{subequations}
\begin{align}
\sfM_{f}^{\setJ}(\bx;n)&= \frac{1}{1+\zeta n} \sum_{w=j}^{j+\zeta n} \E\delta(x_w-t) \label{eq:15c} \\
&= \frac{1}{1+\zeta n} \sum_{w=j}^{j+\zeta n} \rmp^{w(n)}_x(t).
\end{align}
\end{subequations}
As $\zeta\downarrow 0$, $\sfM_{f}^{\setJ}(\bx;n)$ converges to $\rmp^{j(n)}_x(t)$, and thus,
\begin{align}
\rmp^j_x(t)=\lim_{n\uparrow\infty} \lim_{\zeta\downarrow0} \sfM_{f}^{\setJ}(\bx;n). \label{eq:prob_j}
\end{align}
As Proposition \ref{thm:1} indicates, the asymptotic marginal of $f(\bx)$ does not depend on the index set $\setJ(n)$. This concludes that for any $\zeta$ in vicinity of zero, $\lim_{n\uparrow\infty} \sfM_{f}^{\setW}(\bx;n)$ converges to a same value. Consequently, the limit \ac{wrt} $\zeta$ in \eqref{eq:prob_j} can be dropped, and one can write
\begin{align}
\rmp^j_x(t)= \sfM_{f}^{\setJ}(\bx). \label{eq:prob_final}
\end{align}
\eqref{eq:prob_final} indicates that $\rmp^j_x(t)$ is constant in $j$. By substituting \eqref{eq:delta} in Proposition \ref{thm:1}, it is straightforwardly shown that $\sfM_{f}^{\setJ}(\bx)$ is equal to the distribution of $\xx$ at the point $t$, i.e., $\rmp_\xx(t)$, which concludes the proof.
\end{prf}
\begin{remark}
In the proof of Proposition \ref{thm:2}, the convergence of $\sfM_{f}^{\setJ}(\bx;n)$ might be questioned, due to the singularity of the function $f(\cdot)$ defined in \eqref{eq:delta}. To avoid this issue, one can take an alternative approach by defining $f(x)$ to be non-negative integer powers of $x$. In this case, the marginal asymptotic of $f(\bx)$ determines the moments of $x_j$. Using results from the classical moment problem \cite{akhiezer1965classical}, it is then trivial to conclude the decoupling property. More details are found in \cite{guo2005randomly,bereyhi2016rsb}.
\end{remark}
Proposition \ref{thm:2} justifies intuitive findings from Proposition~\ref{thm:1}. In fact, one can consider \eqref{eq:snigle} as a single-user system~whose~expected performance describes the asymptotic average performance of the nonlinear \ac{lse} precoder.
\section{Applications of the Results}
The results given in Section \ref{sec:result} apply to various
\begin{enumerate}[label=(\alph*)]
\item particular precoders as special cases,
\item models of \ac{mimo} fading channels including the well-kn- own \ac{iid} Rayleigh fading model,
\item constraints on \ac{mimo} transmitters, such as peak~and~average power constraints, and \ac{tas}.
\end{enumerate}
In this section, we employ the results to study the asymptotics of some particular precoders. Through out the examples, we consider the channel matrix $\mH$ to be an \ac{iid} fading channel whose entries are \ac{iid} zero-mean random variables~with~variance $n^{-1}$. In this case, $\rmp_\mD$ follows Marcenko-Pastur law \cite{muller2013applications}, and therefore, the $\rmR$-transform reads
\begin{align}
\rmR_\mD(\omega)=\frac{\alpha}{1-\omega}.
\end{align}
\subsection{Average Power Constraint and \ac{tas}}
\label{sec:ex1}
The initial form of the nonlinear \ac{lse} precoder, introduced in \cite{sedaghat2017lse}, considers the penalty function $u(\cdot)$ to be proportional to the Euclidean norm, i.e., $u(\bv)=\lambda\norm{\bv}^2$. This penalty function controls the average power of the precoded vector, i.e., $n^{-1}\E\norm{\bx}^2$, and thus, can satisfy different constraints on the average power by correspondingly setting $\lambda$. Considering the precoder defined in Section \ref{sec:sys}, the generality of the utility function lets the precoder take into account the \ac{tas} in addition to the average power constraint. In order to address both the constraints at the precoder, we consider $u(\cdot)$ to be
\begin{align}
u(\bv)=\lambda \norm{\bv}^2 + \lambda_0 \norm{\bv}_0. \label{eq:penalty0}
\end{align}
The penalty function in \eqref{eq:penalty0} imposes constraints on both the average power and number of active antennas by different values of $\lambda$ and $\lambda_0$ considering a same discussion as given at the beginning of this section. As the first step for investigating the precoder consistent with \eqref{eq:penalty0}, we consider the case in which the precoded symbols are taken from the complex plane, i.e., $\setX=\setC$. In this case, the decoupled precoded symbol reads
\begin{align}
\xx=
\begin{cases}
    \dfrac{s^\rs}{1+\kappa\lambda} \qquad &\abs{s^\rs}\geq \tau \\
    0             &\abs{s^\rs} < \tau \label{eq:sing0}
\end{cases}
\end{align}
where $s^\rs\sim \mathcal{CN}(0, {\lambda^\rs})$ and $\tau\coloneqq\sqrt{\kappa\lambda_0 (1+\kappa\lambda)}$. Here, the variance $\lambda^\rs$, is given by $\lambda^\rs=\alpha^{-1}(\lambda_s + \sfp)$ where $\sfp$ denotes the average transmit power of the precoder, or alternatively, the expected power of the decoupled precoded symbol, i.e.,
\begin{align}
\sfp= \lim_{n\uparrow\infty}\frac{1}{n} \norm{\bx}^2 =\E \abs{\xx}^2= \frac{\lambda^\rs + \tau^2}{\left(1+\kappa\lambda\right)^2} \hspace*{.5mm} e^{-\tfrac{\tau^2}{\lambda^\rs}}. \label{eq:sfp}
\end{align}
Moreover, the scalar $\kappa$ is defined as $\kappa\coloneqq\alpha^{-1}(1 + \chi)$ where the non-negative scalar $\chi$ satisfies
\begin{align}
{\lambda^\rs} \chi&= {\kappa \sfp +\kappa^2 \lambda \sfp}. \label{eq:chi0}
\end{align}
Considering \eqref{eq:sing0}, the decoupled precoded symbol is obtained by hard thresholding the complex Gaussian symbol $s_\rs$ in which the threshold $\tau$ depends on the control factor $\lambda_0$. By setting $\lambda_0=0$, the threshold becomes zero as well, and thus, the decoupled precoded symbol reduces to a complex Gaussian random variable describing the marginal distribution of \ac{rzf} precoder's output symbols. To investigate the constraint on the number of active antennas, we determine the asymptotic fraction of active antennas $\sfa$. Using Proposition \ref{thm:1}, $\sfa$ reads
\begin{align}
\sfa = \lim_{n\uparrow\infty}\frac{1}{n} \norm{\bx}_0=\E \mone \left\lbrace\xx \neq 0 \right\rbrace = e^{-\tfrac{\tau^2}{\lambda^\rs}}. \label{eq:sfa}
\end{align}
where $\mone\set{\cdot}$ denotes the indicator function. Considering given constraints on $\sfp$ and $\sfa$, one can determine the corresponding control factors $\lambda$ and $\lambda_0$, as well as the scalar $\chi$, by solving the set of fixed point equations in \eqref{eq:sfp}, \eqref{eq:chi0} and \eqref{eq:sfa}. The asymptotic distortion is then determined from Proposition~\ref{thm:1}~as
\begin{align}
\sfD=\frac{\lambda_s+\sfp}{\left(1+\chi\right)^2}. \label{eq:sfD0}
\end{align}

\begin{figure}[t]
\hspace*{-1.3cm}  
\resizebox{1.25\linewidth}{!}{
\pstool[width=.35\linewidth]{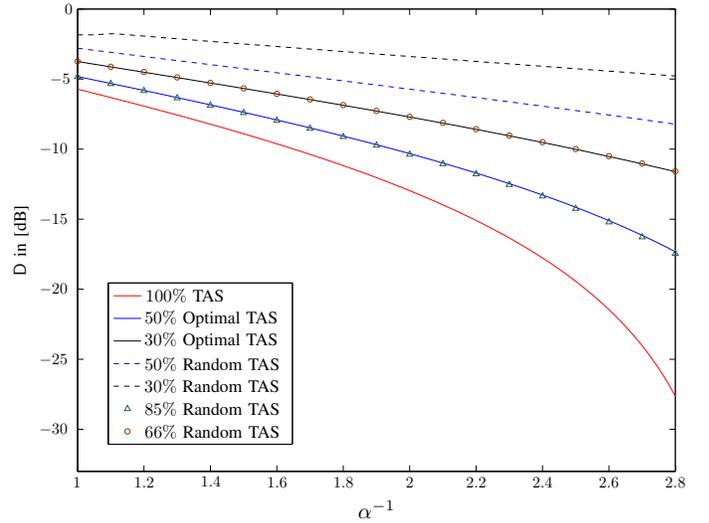}{
\psfrag{Distortion}[c][c][0.2]{$\sfD$ in [dB]}
\psfrag{inverse}[c][c][0.25]{$\alpha^{-1}$}
\psfrag{100-Ant-Select-O}[l][l][0.2]{$100\%$ \ac{tas}}
\psfrag{500-Ant-Select-O}[l][l][0.2]{$50\%$ Optimal \ac{tas}}
\psfrag{300-Ant-Select-O}[l][l][0.2]{$30\%$ Optimal \ac{tas}}
\psfrag{500-Ant-Select-R}[l][l][0.2]{$50\%$ Random \ac{tas}}
\psfrag{300-Ant-Select-R}[l][l][0.2]{$30\%$ Random \ac{tas}}
\psfrag{850-Ant-Select-R}[l][l][0.2]{$85\%$ Random \ac{tas}}
\psfrag{660-Ant-Select-R}[l][l][0.2]{$66\%$ Random \ac{tas}}


\psfrag{-5}[r][c][0.18]{$-5$}
\psfrag{-10}[r][c][0.18]{$-10$}
\psfrag{-15}[r][c][0.18]{$-15$}
\psfrag{-20}[r][c][0.18]{$-20$}
\psfrag{-25}[r][c][0.18]{$-25$}
\psfrag{-30}[r][c][0.18]{$-30$}
\psfrag{0}[r][c][0.18]{$0$}
%
\psfrag{1}[c][b][0.18]{$1$}
\psfrag{1.2}[c][b][0.18]{$1.2$}
\psfrag{1.4}[c][b][0.18]{$1.4$}
\psfrag{1.6}[c][b][0.18]{$1.6$}
\psfrag{1.8}[c][b][0.18]{$1.8$}
\psfrag{2}[c][b][0.18]{$2$}
\psfrag{2.2}[c][b][0.18]{$2.2$}
\psfrag{2.4}[c][b][0.18]{$2.4$}
\psfrag{2.6}[c][b][0.18]{$2.6$}
\psfrag{2.8}[c][b][0.18]{$2.8$}

}}
\caption{\ac{rs} predicted asymptotic distortion as a function of inverse load factor for $\sfp=0.5$ considering \ac{tas} constraints $\sfa=0.5$ and  $\sfa=0.3$. The nonlinear \ac{lse} precoder with penalty function \eqref{eq:penalty0} results in a same distortion as the \ac{rzf} with random \ac{tas} by using about $0.35n$ fewer number of transmit antennas.}
\label{fig:1}
\end{figure}

Fig.~\ref{fig:1} shows the asymptotic distortion in terms of the inverse load factor, i.e., $\alpha^{-1}$, considering various constraints on the fraction of active antennas when the control factor $\lambda$ is selected such that $\sfp=0.5$. Moreover, the source variance is set to be $\lambda_s=1$. As the figure illustrates, the \ac{rs} ansatz predicts that the precoder with a given constraint on $\sfa$ significantly outperforms the \ac{rzf} precoder with random \ac{tas}\footnote{In fact, in this case, the precoder selects a subset of transmit antennas~randomly and precodes $\bs$ using the penalty function $u(\bv)=\lambda\norm{\bv}^2$.}.  To quantify the improvement, we have plotted the asymptotic distortion for the \ac{rzf} precoder with random \ac{tas} considering several values of $\sfa$. The numerical investigations depict that for a given asymptotic distortion, the nonlinear \ac{lse} precoder with penalty function \eqref{eq:penalty0} requires around $0.35n$ fewer number of active antennas compared to a \ac{rzf} precoder with random~\ac{tas}.
\subsection{Peak Power Constraint and \ac{tas}}
\label{sec:ex2}
Active transmit antennas, in practice, are equipped with power amplifiers whose peak powers are restricted. Therefore, assuming the whole complex plane as the set of possible constellation points of the precoder's output is an inaccurate model for many systems. The inaccuracy is more pivotal when the transmit signal is desired to have a relatively small \ac{papr}. This issue can be addressed by enforcing the precoder to take transmit symbols from
\begin{align}
\setX = \set{ r e^{\mathrm{j} \theta }: \hspace*{2mm} 0 \leq \theta \leq 2\pi \ \wedge \ 0 \leq r \leq \sqrt{\rmP}}.
\end{align}
In this case, the output symbols are restricted to lie inside a circle with radius $\sqrt{\rmP}$, and thus, the transmit per-antenna peak power is upper bounded by $\rmP$. By considering a penalty function as in \eqref{eq:penalty0}, we can further impose~a~\ac{tas}~constraint on the precoder. Consequently, the decoupled precoded symbol is given by two steps of hard thresholding as
\begin{align}
\xx=
\begin{cases}
  \dfrac{ s^\rs}{\abs{s^\rs}} \sqrt{\rmP} \qquad  & \hat{\tau} \leq \abs{s^\rs} \\
    0             & \tilde{\tau} \leq \abs{s^\rs} < \hat{\tau} \\
    \dfrac{s^\rs}{1+\kappa\lambda} &\tau \leq \abs{s^\rs}\leq \tilde{\tau} \\
    0             &0\leq \abs{s^\rs} < \tau \label{eq:single1}
\end{cases}
\end{align}
where $s^\rs\sim \mathcal{CN}(0, {\lambda^\rs})$ and the thresholds $\tau$, $\tilde{\tau}$, and $\hat{\tau}$ read
\begin{subequations}
\begin{align}
\tau&\coloneqq\sqrt{\kappa \lambda_0 (1+\kappa\lambda)} \label{eq:tau} \\
\tilde{\tau}&\coloneqq (1+\kappa\lambda)\sqrt{\rmP} \label{eq:tau_t} \\
\hat{\tau}&\coloneqq\max \set{ (1+\kappa\lambda)\sqrt{\rmP},\frac{1+\kappa\lambda}{2} \sqrt{\rmP} +\frac{\kappa\lambda_0}{2\sqrt{\rmP}}} \label{eq:tau_h}
\end{align}
\end{subequations}
The variance $\lambda^\rs$, is determined as $\lambda^\rs=\alpha^{-1}(\lambda_s + \sfp)$ with
\begin{align}
\sfp= \lim_{n\uparrow\infty}\frac{1}{n} \norm{\bx}^2 =\E \abs{\xx}^2= \frac{\Xi}{(1+\kappa\lambda)^2} + \rmP \hspace*{.5mm} e^{-\tfrac{\hat{\tau}^2}{\lambda^\rs}}. \label{eq:p_1}
\end{align}
being the average transmit power of the precoder where
\begin{align}
\Xi=(\lambda^\rs+\tau^2)\hspace*{.5mm} e^{-\tfrac{\tau^2}{\lambda^\rs}}-(\lambda^\rs+\tilde{\tau}^2)\hspace*{.5mm}e^{-\tfrac{\tilde{\tau}^2}{\lambda^\rs}}.
\end{align}
Moreover, $\kappa\coloneqq\alpha^{-1}(1 + \chi)$ where $\chi$ satisfies
\begin{align}
\hspace*{-2mm}\chi= \frac{\kappa \hspace*{.3mm}\Xi}{\lambda^\rs(1+\kappa \lambda)}+ \frac{\kappa\hat{\tau}\sqrt{\rmP}}{\lambda^\rs}  \hspace*{.5mm} e^{-\tfrac{\hat{\tau}^2}{\lambda^\rs}} +  \kappa\sqrt{\frac{\pi \rmP}{\lambda^\rs}}\hspace*{.5mm}\rmQ(\sqrt{\frac{2}{\lambda^\rs}}\hat{\tau}). \label{eq:chi_1}
\end{align}
with $\rmQ\left( \cdot \right)$ representing the standard $\rmQ$-function. Considering \eqref{eq:single1}, the decoupled precoded symbol is obtained from $s^\rs$ by a two steps hard thresholding. In the first step, $s^\rs$ is compared to $\tilde{\tau}$, in order to be constrained \ac{wrt} the peak power $\rmP$. The second step, then, imposes the \ac{tas} constraint on the precoded symbol using the thresholds $\tau$ and $\hat{\tau}$. As a result, $\tau$ and $\hat{\tau}$ depend on the control factor $\lambda_0$. By setting $\lambda_0=0$, we have $\tau=0$ and $\hat{\tau}=\tilde{\tau}$, and thus, the precoder reduces to the \ac{papr} limited precoder studied in \cite{sedaghat2017lse}. To investigate the \ac{tas} and \ac{papr} constraints imposed on the precoder, we further determine the asymptotic fraction of the active antennas, as well as the transmit signal's \ac{papr}. Using Proposition \ref{thm:1}, the asymptotic fraction of active antennas is given by
\begin{align}
\hspace*{-2mm}\sfa \hspace*{-.7mm}=\hspace*{-.7mm} \lim_{n\uparrow\infty}\frac{1}{n} \norm{\bx}_0 \hspace*{-.7mm} =\hspace*{-.7mm} \E \mone \set{\xx\neq 0} \hspace*{-.7mm}= \hspace*{-.7mm} e^{-\tfrac{\tau^2}{\lambda^\rs}}+e^{-\tfrac{\hat{\tau}^2}{\lambda^\rs}}-e^{-\tfrac{\tilde{\tau}^2}{\lambda^\rs}}. \label{eq:a_1}
\end{align}
Moreover, considering the average transmit power $\sfp$ in \eqref{eq:p_1}, the asymptotic \ac{papr} of the precoder is $\papr={\sfp}^{-1}{\rmP}$.~Consequently, for given constraints on $\sfa$ and $\papr$, the control factors $\lambda$ and $\lambda_0$ and the scalar $\chi$ are found through the fixed point equations \eqref{eq:p_1}, \eqref{eq:chi_1} and \eqref{eq:a_1}, and the asymptotic distortion is determined as in \eqref{eq:sfD0}.

\begin{figure}[t]
\hspace*{-1.3cm}  
\resizebox{1.25\linewidth}{!}{
\pstool[width=.35\linewidth]{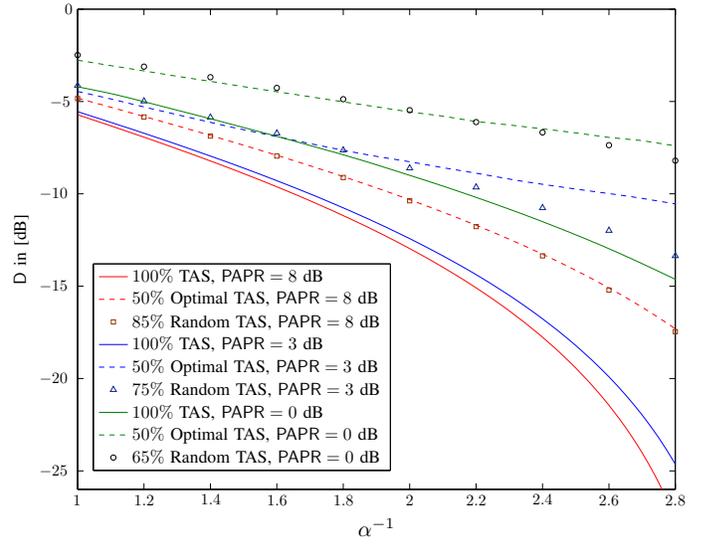}{
\psfrag{Distortion}[c][c][0.2]{$\sfD$ in [dB]}
\psfrag{inverse}[c][c][0.25]{$\alpha^{-1}$}
\psfrag{PAPR=3dB-100-Ant-Select-O}[l][l][0.2]{$100\%$ \ac{tas}, $\papr=3$ dB}
\psfrag{PAPR=8dB-100-Ant-Select-O}[l][l][0.2]{$100\%$ \ac{tas}, $\papr=8$ dB}
\psfrag{PAPR=0dB-100-Ant-Select-O}[l][l][0.2]{$100\%$ \ac{tas}, $\papr=0$ dB}
\psfrag{PAPR=0dB-500-Ant-Select-O}[l][l][0.2]{$50\%$ Optimal \ac{tas}, $\papr=0$ dB}
\psfrag{PAPR=3dB-500-Ant-Select-O}[l][l][0.2]{$50\%$ Optimal \ac{tas}, $\papr=3$ dB}
\psfrag{PAPR=8dB-500-Ant-Select-O}[l][l][0.2]{$50\%$ Optimal \ac{tas}, $\papr=8$ dB}
\psfrag{PAPR=0dB-650-Ant-Select-R}[l][l][0.2]{$65\%$ Random \ac{tas}, $\papr=0$ dB}
\psfrag{PAPR=3dB-750-Ant-Select-R}[l][l][0.2]{$75\%$ Random \ac{tas}, $\papr=3$ dB}
\psfrag{PAPR=8dB-850-Ant-Select-R}[l][l][0.2]{$85\%$ Random \ac{tas}, $\papr=8$ dB}


\psfrag{-5}[r][c][0.18]{$-5$}
\psfrag{-10}[r][c][0.18]{$-10$}
\psfrag{-15}[r][c][0.18]{$-15$}
\psfrag{-20}[r][c][0.18]{$-20$}
\psfrag{-25}[r][c][0.18]{$-25$}
\psfrag{-30}[r][c][0.18]{$-30$}
\psfrag{0}[r][c][0.18]{$0$}
%
\psfrag{1}[c][b][0.18]{$1$}
\psfrag{1.2}[c][b][0.18]{$1.2$}
\psfrag{1.4}[c][b][0.18]{$1.4$}
\psfrag{1.6}[c][b][0.18]{$1.6$}
\psfrag{1.8}[c][b][0.18]{$1.8$}
\psfrag{2}[c][b][0.18]{$2$}
\psfrag{2.2}[c][b][0.18]{$2.2$}
\psfrag{2.4}[c][b][0.18]{$2.4$}
\psfrag{2.6}[c][b][0.18]{$2.6$}
\psfrag{2.8}[c][b][0.18]{$2.8$}

}}
\caption{Asymptotic distortion versus the inverse load factor for various constraints on the \ac{papr} and $\sfa$ considering $\sfp=0.5$ and $\lambda_s=1$. For $\papr=8$ dB, the precoder performs approximately same as the case with no peak power constraint, and thus, the gain obtained by the penalty function \eqref{eq:penalty0} is same as Fig~\ref{fig:1}. The gain, however, reduces for smaller \ac{papr}s.}
\label{fig:2}
\end{figure}

In Fig.~\ref{fig:2}, the asymptotic distortion is given as a function of the inverse load factor for various constraints on the \ac{papr} and number of active antennas considering $\sfp=0.5$ and~${\lambda_s=1}$. For sake of comparison, we have also plotted the curves for \ac{papr} limited precoders with random \ac{tas} fitted numerically to the curves of \ac{papr} limited precoders with \ac{tas} constraint. As the figure depicts, at higher \ac{papr}s, e.g., $\papr=8$ dB, the precoder outperforms random \ac{tas} with an approximately same reduction in the fraction of active antennas as for the case considered in Section \ref{sec:ex1} (around $0.35n$). The result was also reported in \cite{sedaghat2017lse} where the authors showed that the performance of \ac{papr} limited precoders converges to the case with no peak power constraint in relatively small \ac{papr}s. As the \ac{papr} decreases to $0$ dB, the reduction in the fraction of active antennas decreases as well. For example, as Fig.~\ref{fig:2} illustrates, considering $\papr=3$~dB, the reduction is around $0.25n$ for load factors near to $1$ while for the constant envelope case, i.e., $\papr=0$ dB, it reduces to $0.15n$.

\section{Conclusion}
A general nonlinear \ac{lse} precoding scheme for the downlink of massive \ac{mimo} channels has been considered. The scheme addresses several hardware limitations on the transmitter including limited \ac{papr} and number of active transmit antennas. Using the replica method from statistical mechanics, we have investigated the large-system performance of the precoder under the \ac{rs} assumption. The numerical investigations have shown that for a given distortion at the receiver side, the proposed precoder needs $35\%$ fewer active transmit antennas compared to the random \ac{tas}. An underway~extension~of~this work is to consider \ac{rsb} ans\"atze, in order to determine~more~accurate~predictions~of~the performance in the regimes that \ac{rs} fails.

\appendix
\section*{Sketch of the Proof}
\label{sec:large}
In this section, we explain the strategy for deriving Proposition \ref{thm:1}. The proof is based on the replica method developed in statistical mechanics. The detailed derivations are omitted here and left for the extended version of the manuscript. To start with, consider the function
\begin{align}
\mae(\bv|\bs,\mH)=\norm{\mH \bv-\bs}^2 +u(\bv), \label{eq:8}
\end{align}
being referred to as the ``Hamiltonian''. We define the partition function $\maz(\beta,h)$ to be
\begin{align}
\maz(\beta,h) =\sum_{\bv} e^{-\beta\mae(\bv|\bs,\mH)+hn\sfM^{\setW}_f(\bv;n)}.
\end{align}
\subsection{Evaluation of the Asymptotic Marginal}
\label{ssec:asy_marg}
Using the standard large deviation argument, it is shown that the asymptotic marginal of $f(\bx)$ reads
\begin{align}
\sfM^{\setW}_f(\bx) =\lim_{n\uparrow\infty}  \lim_{\beta\uparrow\infty}  \frac{\partial}{\partial h} \maf(\beta,h)|_{h=0}, \label{eq:10}
\end{align}
where we have defined the function $\maf(\cdot)$ to be
\begin{align}
\maf(\beta,h) \coloneqq \frac{1}{n}  \E \log \maz(\beta,h).
\end{align}
Therefore, the evaluation of the asymptotic marginal reduces to determination of the function $\maf(\cdot)$. We further show that the asymptotic distortion is directly determined from $\maf(\cdot)$.
\subsection{Evaluation of the Asymptotic Distortion}
\label{ssec:asy_dist}
Considering the Hamiltonian in \eqref{eq:8}, it can be shown that the asymptotic distortion satisfies
\begin{align}
\alpha \sfD + \sfM_u^\setT (\bx) = \tilde{\mae} \label{eq:13}
\end{align}
where $\sfM_u^\setT (\bx)$ denotes the asymptotic marginal of $u(\bx)$ over $\setT(n) \coloneqq [1:n]$, and $\tilde{\mae}$ is the asymptotic average energy of the Hamiltonian\footnote{In the context of statistical mechanics, $\tilde{\mae}$ is the average energy of the spin glass defined by the Hamiltonian \eqref{eq:8} in the thermodynamic limit.} defined as
\begin{align}
\tilde{\mae}={\lim_{n\uparrow\infty} \frac{1}{n} \E \mae(\bx|\bs,\mH)}.
\end{align}
Using \eqref{eq:10}, $\sfM_u^\setT (\bx)$ is determined in terms of $\maf(\cdot)$. Moreover, 
\begin{align}
\tilde{\mae} = - \lim_{n\uparrow\infty}  \lim_{\beta\uparrow\infty}  \frac{\partial}{\partial \beta} \maf(\beta,h)|_{h=0}. \label{eq:14}
\end{align}
Therefore, the asymptotic distortion $\sfD$ can be evaluated from $\maf(\cdot)$ considering the equality in \eqref{eq:13}.
\subsection{Determining $\maf(\cdot)$}
Considering the discussions in Sections \ref{ssec:asy_marg} and \ref{ssec:asy_dist}, the large-system analysis of the precoder reduces to determining the function $\maf(\cdot)$. We do this task by employing the replica method which has been initially developed in statistical mechanics for analysis of spin glasses \cite{edwards1975theory} and later employed in information theory to investigate the asymptotics of different problems; see \cite{tanaka2002statistical,guo2005randomly, rangan2012asymptotic, bereyhi2016rsb} and the references therein.

To determine $\maf(\cdot)$, one needs to overcome the hard task of taking expectation of a logarithmic function. The task becomes analytically non-traceable when the argument of the logarithm is sum of exponential functions. Alternatively, one can employ the Riesz equality which states that for a random variable $\xx$ 
\begin{align}
\E \log \xx = \lim_{m\downarrow 0} \frac{1}{m} \log \E \xx^m,
\end{align}
and bypasses the logarithmic expectation by writing $\maf(\cdot)$ as 
\begin{align}
\maf(\beta,h) = \frac{1}{n} \lim_{m\downarrow 0}  \frac{1}{m} \log \E \left[ \maz(\beta,h) \right]^m. \label{eq:18}
\end{align}
\subsection{Replica Method}
The computation of \eqref{eq:18} is still non-trivial, since the \ac{rhs} of \eqref{eq:18} needs to be determined for real values of $m$ (or at least, for some $m$ in a right neighborhood of $0$). Replica method determines \eqref{eq:18} by considering the conjecture of the replica continuity. The replica continuity indicates that the analytic continuation of the non-negative integer moment function, i.e., $\E \left[ \maz(\beta,h) \right]^m$ for $m\in\setZ^+$, onto the set of non-negative real numbers equals to the non-negative real moment function, i.e., $\E \left[ \maz(\beta,h) \right]^m$ for $m\in\setR^+_0$. More intuitively, it suggests us to determine the moment function as a function in $m\in\setZ^+$, and then, assume that the function is of the same form for $m\in\setR_0^+$. The rigorous justification of the replica continuity has not been yet precisely addressed; however, the analytic results from the theory of spin glasses confirm the validity of the conjecture for several~cases.~Following~the~replica continuity, the moment function $\sfM(m)\coloneqq\E \left[ \maz(\beta,h) \right]^m$ reads
\begin{subequations}
\begin{align}
\sfM(m) &\coloneqq \E \sum_{\set{\bv_a}} \prod_{a=1}^m e^{-\beta\mae(\bv_a|\bs,\mH)+hn\sfM^{\setW}_f(\bv_a;n)} \\
&= \sum_{\set{\bv_a}} \left[ \E e^{-\beta\sum\limits_{a=1}^m \mae(\bv_a|\bs,\mH)} \right] e^{hn\sum\limits_{a=1}^m \sfM^{\setW}_f(\bv_a;n)}
\end{align}
\end{subequations}
where $\set{\bv_a}\coloneqq\set{\bv_1, \cdots, \bv_m}$. Taking the expectation \ac{wrt} $\bs$, the moment function reduces to
\begin{align}
\sfM(m)= \sum_{\set{\bv_a}} \E_{\mJ} e^{-\beta\sum\limits_{a,b=1}^m \bv_a^\her \mJ \bv_b \xi_{ab} - n \Theta \set{\bv_a}} \label{eq:20}
\end{align}
where $\xi_{ab}\coloneqq \delta(a-b) - \lambda_s \beta (1+\lambda_s \beta m)^{-1}$, $\mJ$ denotes the Gramian of $\mH$, i.e. $\mJ\coloneqq \mH^\her \mH$, and
\begin{align}
\Theta \set{\bv_a} \coloneqq \frac{1}{n} \sum_{a=1}^m \left[ \beta u(\bv_a)- hn\sfM^{\setW}_f(\bv_a;n)\right] + \frac{1}{n}\Delta_m 
\end{align}
with $\Delta_m\coloneqq k \log (1+\lambda_s\beta m)$. In order to take the expectation \ac{wrt} $\mJ$, we invoke the result reported in \cite{guionnet2005fourier} for spherical integrals, where a closed form formula has been given under a set of assumptions. At this point, let us define $\mQ_m $ to be an $m\times m$ matrix with entries
\begin{align}
[\mQ_m]_{ab}=\frac{1}{n} \bv_a^\her \bv_b. \label{eq:Q}
\end{align}
Then, after taking the expectation \ac{wrt} $\mJ$, \eqref{eq:20} reduces to
\begin{align}
\sfM(m)= \sum_{\set{\bv_a}} e^{-n\mg (\mT \mQ_m) - n \Theta \set{\bv_a}} \label{eq:23}
\end{align}
where $\mT \coloneqq \mI_m- \lambda_s \beta (1+\lambda_s \beta m)^{-1} \mone_m$, and $\mg(\cdot)$ reads
\begin{align}
\mg (\mM) &= \sum_{\ell=1}^m \int_0^{\beta \lambda_\ell} \rmR_{\mD} (-\omega) \dif \omega \label{eq:24}
\end{align}
for some matrix $\mM_{m\times m}$ with eigenvalues $\set{\lambda_\ell}$ for $\ell\in[1:m]$. The function $\rmR_\mD (\cdot)$ in \eqref{eq:24} denotes the $\rmR$-transform of the asymptotic distribution $\rmp_\mD$ defined in Definition \ref{def:r-trans}. The sum in \eqref{eq:23} can be determined by dividing the set of vectors $\set{\bv_a}$ into subshells regarding $\mQ_m$. More precisely, define the subshell $\mas(\mQ)$ as the set of vectors $\set{\bv_a}$ in which the corresponding matrix $\mQ_m$ equals to $\mQ$; then, \eqref{eq:23} can be written as
\begin{align}
\sfM(m)= \int e^{-n\mg (\mT \mQ) - n\mai(\mQ)} \dif \mQ \label{eq:25}
\end{align}
where $\dif \mQ \coloneqq \prod_{a,b=1}^m \dif \re{[\mQ]_{ab}} \dif \img{[\mQ]_{ab}}$, the integral is taken over $\setC^{m\times m}$, and $e^{- n\mai(\mQ)}$ indicates the density of ${\mas(\mQ)}$ which is written as
\begin{align}
e^{-n\mai(\mQ)}=\sum_{\set{\bv_a}} e^{ - n \Theta \set{\bv_a}} w(\mQ;\set{\bv_a}) \label{eq:26}
\end{align}
with the weight function $w(\mQ;\set{\bv_a})$ being
\begin{align}
w(\mQ;\set{\bv_a})=\prod_{a,b=1}^m &\delta(\re{n[\mQ]_{ab}-\bv_a^\her \bv_b}) \nonumber \\ &\times \delta(\img{n [\mQ]_{ab} -\bv_a^\her \bv_b}). \label{eq:27}
\end{align}
The weight function \eqref{eq:27} is further calculated in an analytic form using the Laplace inverse transform of the~impulse~function $\delta(\cdot)$. 

At this point, $\maf(\cdot)$ is determined by substituting \eqref{eq:25} in~\eqref{eq:18} and taking the limits. We assume that the limits \ac{wrt} $n$ and $m$ commutate. This assumption is common in replica analyses and is intuitively concluded from the analytic continuity of the moment function being conjectured by the replica continuity. Consequently, $n$ in \eqref{eq:25} can be considered to grow large, and thus, by the saddle point method, one can conclude that
\begin{align}
\sfM(m) \doteq \sfK_n  e^{-n\left[ \mg (\mT \tilde{\mQ}) +  \mai(\tilde{\mQ})\right]} \label{eq:30}
\end{align}
where $\doteq$ denotes the asymptotic logarithmic equivalence\footnote{$a_n \doteq b_n$ if $\lim_{n\uparrow\infty}\abs{\frac{a_n}{b_n}}=0$.}, $\sfK_n$ is a non-negative real sequence in $n$ bounded for any $n$, and $\tilde{\mQ}$ is the saddle point of the exponent function. 
\subsection{\ac{rs} assumption}
The next step is to determine the saddle point $\tilde{\mQ}$, explicitly by searching over all possible $m\times m$ matrices consistent with \eqref{eq:Q}. The task, however, is not feasible, due to the both complexity and analyticity issues. In fact, one may find the above task to be computationally complex. Moreover, some of the solutions for the saddle point may result in non-analytic moment functions which cannot be analytically continued to the real axis, and thus, are not of use. To address both the issues, we invoke the well-known trick from theory of spin glasses which suggests to restrict the search to a set of parameterized matrices. The known structures for these sets are conjectured from the physical intuition behind the replica analysis and are mainly \ac{rs} or \ac{rsb}. In \ac{rs}, the replicas, i.e., $\set{\bv_a}$, are assumed to behave symmetrically at the saddle point meaning that interactions between any two replicas $\bv_a$ and $\bv_b$ are the same. Therefore, due to \eqref{eq:Q}, $\tilde{\mQ}$ in this case is invariant under all permutations, meaning that ${\mP}^{-1} \mQ {\mP}=\mQ$ for all permutation matrices ${\mP}$ taken from the symmetric group on $[1:m]$. This assumption leads to the following structure
\begin{align}
\tilde{\mQ}= \frac{\chi}{\beta} \mI_m + \sfp \mone_m \label{eq:rs}
\end{align}
for some non-negative real scalars $\chi$ and $q$. For various problems, the \ac{rs} structure results in solutions which are consistent with the available rigorous bounds \cite{guo2005randomly,rangan2012asymptotic,tanaka2002statistical,muller2008vector}. There are, however, several examples in which the \ac{rs} ansatz does not give valid solutions \cite{sedaghat2017lse,bereyhi2016rsb,zaidel2012vector,bereyhi2016statistical}. For these cases, it is believed that the problem is caused by the assumed structure,~i.e.,~\ac{rs}, and therefore, more general structures are supposed to result in valid solutions. In \cite{parisi1980sequence}, Parisi proposed the iterative method of \ac{rsb} which extends \ac{rs} to a sequence of more general structures. The authors in \cite{zaidel2012vector, bereyhi2016rsb, sedaghat2017lse}, employed Parisi's method to investigate the examples with invalid \ac{rs} ansatz considering \ac{rsb} ans\"atze. We believe that the examples in Sections \ref{sec:ex1} and \ref{sec:ex2}, at least for the range of system parameters which we consider, are investigated by the \ac{rs} ansatz considerably accurate. Therefore, at this point we consider $\tilde{\mQ}$ to be of~the~form \eqref{eq:rs}, and leave the derivation of \ac{rsb} ans\"atze as a future~work. After some lines of derivations, the scalars $\chi$ and $\sfp$ at the saddle point are found as
\begin{subequations}
\begin{align}
\sfp&=\int \abs{\arg\min_v \mae_\rs(v|s_0)}^2 \hspace*{.5mm} \md s_0, \label{eq:46a} \\
\chi&= \frac{1}{\varsigma}\int \re{\arg\min_v \mae_\rs(v|s_0) \hspace*{.8mm} s_0^*} \hspace*{.5mm} \md s_0, \label{eq:46b}
\end{align}
\end{subequations}
where $\mae_\rs(\cdot|s_0):\setX\mapsto\setR$ for a given $s_0\in\setC$ is defined as
\begin{align}
\mae_\rs(v|s_0)\coloneqq \abs{v-\dfrac{\varsigma}{\rmR_\mD(-\chi)} \hspace*{.5mm} s_0}^2 + \dfrac{u(v)}{\rmR_\mD(-\chi)}, \label{eq:e_rs}
\end{align}
$\md s_0 \coloneqq \dfrac{1}{\pi} e^{-\abs{s_0}^2} {\dif \re{s_0} \dif \img{s_0} }$, and $\varsigma$ is given by
\begin{align}
\varsigma^2=\frac{\partial}{\partial \chi} \left[ ( \lambda_s \chi- \sfp ) \rmR_\mD(-\chi) \right].
\end{align}
The function $\maf (\cdot)$ is then determined by substituting $\tilde{\mQ}$ with $\chi$ and $\sfp$ given in \eqref{eq:46a} and \eqref{eq:46b}. Considering the discussions in Sections \ref{ssec:asy_marg} and \ref{ssec:asy_dist}, the asymptotic marginal of $f(\bx)$ is calculated as
\begin{align}
\sfM^{\setW}_f(\bx)=  \int f\left(\arg\min_v \mae_\rs\left(v|s_0\right) \right) \md s_0,
\end{align}
and the asymptotic distortion is determined by
\begin{align}
\sfD=\lambda_s+\alpha^{-1} \frac{\partial}{\partial \chi} \left[ ( \sfp -\lambda_s \chi ) \chi \rmR_\mD(-\chi) \right]
\end{align}

Defining the scalar $\lambda^\rs=\varsigma \left[\rmR_\mD(-\chi)\right]^{-1}$, Proposition~\ref{thm:1} is finally concluded. Detailed derivations of this section are given in the extended version of the manuscript.

\bibliography{ref}
\bibliographystyle{IEEEtran}

\begin{acronym}
\acro{mimo}[MIMO]{Multiple-Input Multiple-Output}
\acro{csi}[CSI]{Channel State Information}
\acro{awgn}[AWGN]{Additive White Gaussian Noise}
\acro{iid}[i.i.d.]{independent and identically distributed}
\acro{ut}[UT]{User Terminal}
\acro{bs}[BS]{Base Station}
\acro{tas}[TAS]{Transmit Antenna Selection}
\acro{lse}[LSE]{Least Square Error}
\acro{rhs}[r.h.s.]{right hand side}
\acro{lhs}[l.h.s.]{left hand side}
\acro{wrt}[w.r.t.]{with respect to}
\acro{rs}[RS]{Replica Symmetry}
\acro{rsb}[RSB]{Replica Symmetry Breaking}
\acro{papr}[PAPR]{Peak-to-Average Power Ratio}
\acro{rzf}[RZF]{Regularized Zero Forcing}
\acro{snr}[SNR]{Signal-to-Noise Ratio}
\acro{rf}[RF]{Radio Frequency}
\acro{mf}[MF]{Match Filtering}
\end{acronym}

\end{document}